\documentclass{emulateapj} 
\usepackage{apjfonts} 
\usepackage{amsmath}
\usepackage{amssymb}
\usepackage{color}
\usepackage{natbib}

\shorttitle{Velocity Dispersions and Shapes of Halo-Star Orbits} \shortauthors{Hattori, Yoshii, Beers, Carollo, \& Lee}

\begin{document}

\title{Very Metal-Poor Outer-Halo Stars with Round Orbits} 

\author{
Kohei Hattori\altaffilmark{1}, 
Yuzuru Yoshii\altaffilmark{1}, 
Timothy C. Beers\altaffilmark{2,3},  
Daniela Carollo\altaffilmark{4,5}, and 
Young Sun Lee\altaffilmark{6}}

\altaffiltext{1}{Institute of Astronomy, School of Science, University of Tokyo, 2-21-1, Osawa, Mitaka, Tokyo 181-0015, Japan}
\altaffiltext{2}{National Optical Astronomy Observatories, Tucson, AZ 85719, USA}
\altaffiltext{3}{Department of Physics \& Astronomy and JINA: Joint Institute for Nuclear Astrophysics, Michigan State University, E. Lansing, MI 48824, USA}
\altaffiltext{4}{Macquarie University - Dept. Physics \& Astronomy, Sydney, 2109 NSW, Australia}
\altaffiltext{5}{INAF - Osservatorio Astronomico di Torino, 10025 Pino Torinese, Torino - Italy}
\altaffiltext{6}{Department of Astronomy, New Mexico State University, Las Cruces, NM 88003, USA}
\email{khattori@ioa.s.u-tokyo.ac.jp}

\begin{abstract}
The orbital motions of halo stars in the Milky Way reflect the orbital motions of the progenitor systems in which they formed, making it possible to trace the mass-assembly history of the Galaxy. Direct measurement of three-dimensional velocities, based on accurate proper motions and line-of-sight velocities, has revealed that the majority of halo stars in the inner-halo region move on eccentric orbits. However, our understanding of the motions of distant, in-situ halo-star samples is still limited, due to the lack of accurate proper motions for these stars. Here we explore a model-independent analysis of the line-of-sight velocities and spatial distribution of a recent sample of 1865 carefully selected halo blue horizontal-branch (BHB) stars within $30\;{\rm kpc}$ of the Galactic center. We find that the mean rotational velocity of the very metal-poor (${\rm[Fe/H]}< -2.0$) BHB stars significantly lags behind that of the relatively more metal-rich (${\rm[Fe/H]}> -2.0$) BHB stars. We also find that the relatively more metal-rich BHB stars are dominated by stars with eccentric orbits, as previously observed for other stellar samples in the inner-halo region. By contrast, the very metal-poor BHB stars are dominated by stars on rounder, lower-eccentricity orbits. Our results indicate that the motion of the progenitor systems of the Milky Way that contributed to the stellar populations found within $30\;{\rm kpc}$ correlates directly with their metal abundance, which may be related to their physical properties such as gas fractions. These results are consistent with the existence of an inner/outer halo structure for the halo system, as advocated by \citeauthor{Carollo2010} 
\end{abstract}


\section{Introduction}

The luminosity of the Milky Way is dominated by its disk, where the
great majority of stars ($>90\%$) are found. By comparison, out of $\sim
10^{11}$ stars in the Milky Way, the stellar halo comprises only a tiny
fraction ($\sim 1\%$), but this component is a precious source of
information on the formation history of the Galaxy. First, halo stars
are very old ($\sim 10-13\;{\rm Gyrs}$), and their chemical compositions
provide information on the ancient environments in which these stars
formed. Typically, the metal abundances of halo stars are less than
1/10th of the Solar value \citep{CB2000, Carollo2007, Carollo2010},
which immediately suggests that the metal enrichment due to supernova
explosions in the Universe had not progressed very far when halo stars
were formed. Secondly, the stellar halo is a collisionless system, hence
two-body relaxation is expected to be unable to fully erase the
initial orbital properties of halo stars. It follows that the present
motions of halo stars reflects their motions in the early Universe,
which can be used to explore the kinematics of their progenitor systems,
such as the gas clouds or dwarf galaxies in which these halo stars
formed. 
 
To date, detailed analyses of the chemical and dynamical properties of
halo stars has been confined to stars up to $\sim 10-15\;{\rm kpc}$ from the
Sun (centered at the Galactocentric distance of the Sun, $\sim 8.5\;{\rm
kpc}$), although the spatial distribution of the stellar halo extends to
$\sim 100\;{\rm kpc}$ or more. This is mainly because we do not possess
sufficiently accurate measurements of proper motions for more distant
halo stars. In order to avoid this limitation, many authors have studied
the line-of-sight velocities and spatial distribution of halo stars,
making use of various kinematic models -- such as distribution function
models of the stellar halo, or gravitational potential models of the
Milky Way. However, the conclusions of previous studies on the orbital
distribution of halo stars well outside the local region are divergent.
Some suggest tangentially-anisotropic orbital distributions
\citep{SL1997, Kafle2012}, others suggest radially-anisotropic
distributions \citep{Deason2012}, and still others suggest nearly
isotropic distributions \citep{Sirko2004, Thom2005}. This might imply
that the stellar halo is not a simple entity which can be described by a
single distribution function model. Indeed, based on observations of
relative nearby ($d \le 4\;{\rm kpc}$) halo stars, \citep{Carollo2007,
Carollo2010} suggest that a dual-halo model is more appropriate, in
which the stellar halo consists of a relatively metal-rich inner-halo
component with a net zero to slightly prograde rotation, and a very
metal-poor outer-halo component with a net retrograde rotation. Recent
observations of retrograde outer-halo RR Lyrae stars by
\cite{Kinman2012}, and a model-fitting analysis for distant halo stars
by \cite{Deason2011} supports this idea, as do recent numerical
simulations of the formation of Milky Way-like galaxies (e.g.,
\citealt{McCarthy2012}). Here we introduce a new analysis of the halo
system, which requires only a minimum of assumptions, and does not
require any kinematic models. 

This paper is outlined as follows. In Section 2 we describe our sample
selection and expected errors in distance, line-of-sight velocities, and
metallicities. Section 3 describes our analysis approach. In Section 4, we
examine the application of this technique to our sample of halo BHB
stars, segregated on metallicity. Section 5 presents a brief discussion
and conclusions.  
 
\section{Sample Selection}

Our sample comprises 1865 blue horizontal-branch (BHB) stars from 
Data Release 8 of the Sloan Digital Sky Survey \citep{Aihara2011}, with
Galactocentric distances in the range $6<r/{\rm kpc}<30$, as originally
selected and carefully validated by \cite{Xue2011}. This sample is free
from significant contamination by the Sagittarius stream and the thick
disk, as we apply a spatial masking scheme in their selection. Distances
from the Sun are accurate to $\sim5-10\%$, and the line-of-sight
velocity errors are $5-20\;{\rm km\;s^{-1}}$ (see \citealt{Xue2011}).
Stellar metallicities, [Fe/H], for this sample are also available.
Following \cite{Xue2011}, we use the metallicities obtained by the
\citeauthor{Wilhelm1999} methodology in the SEGUE Stellar Parameter
Pipeline (SSPP; see \citealt{Lee2008} for details), which are likely to
be the most reliable for stars with effective temperatures of BHB stars
(on the order of 0.3 dex). We then divide this sample on metallicity --
994 of our stars are relatively metal-rich (${\rm [Fe/H]}>-2.0$), while
871 stars are very metal-poor (${\rm[Fe/H]} < -2.0$).\footnote{
Our boundary metallicity (${\rm[Fe/H]} = -2.0$) lies between the peak
metallicities of the inner-halo (${\rm[Fe/H]_{peak}} \simeq -1.6$) and
outer-halo (${\rm[Fe/H]_{peak}} \simeq -2.2$) components
\citep{Carollo2007}. }

\section{Analysis Method}

\subsection{Derivation of Rotational Velocity}

Let us denote by $S$ an imaginary observer located at the Sun who is at
rest with respect to the Galactic rest frame, and by $O$ an observer
located at the Sun who moves with the Sun. Here, we assume that the
velocity of $O$ with respect to $S$, $\textbf{v}_\odot$, and the
three-dimensional (3D) position of the Sun with respect to the Galactic
center are known. 

Now, suppose that the $k$-th star ($k=1,\cdots, n$) is observed in the
direction of $\textbf{x}^{los}_k$ by $S$ and $O$.
Then, the line-of-sight velocity with respect to $S$
can be expressed as: 

\begin{equation} 
	v^{los}_k \equiv \textbf{v}_k \cdot \textbf{x}^{los}_k , 
\end{equation}

\noindent where $\textbf{v}_k$ is the velocity of this star with respect to $S$. 
Since the line-of-sight velocity of this star with respect to $O$ is: 

\begin{equation} \label{v_hel}
	v^{los, hel}_k \equiv (\textbf{v}_k - \textbf{v}_\odot) \cdot \textbf{x}^{los}_k = v^{los}_k - \textbf{v}_\odot \cdot \textbf{x}^{los}_k , 
\end{equation} 

\noindent we can calculate $v^{los}_k$ from ($\textbf{v}_\odot, v^{los, hel}_k, \textbf{x}^{los}_k$). 

On the other hand, if we decompose $\textbf{v}_k$ as: 

\begin{equation}
	\textbf{v}_{k} = v_{r,k} \textbf{e}_{r,k} + v_{\theta,k} \textbf{e}_{\theta,k} + v_{\phi,k} \textbf{e}_{\phi,k} , 
\end{equation}

\noindent we obtain 

\begin{equation} \label{v_s}
	v^{los}_k = v_{r, k} Q_{r,k} + v_{\theta, k} Q_{\theta,k} + v_{\phi,k} Q_{\phi,k} , 
\end{equation}

\noindent where 

\begin{equation} 
	Q_{i,k} \equiv \textbf{x}^{los}_k \cdot \textbf{e}_{i,k} \;(i=r,
\theta,\phi) .
\end{equation} 

\noindent Here, $v_{i,k}$ and $\textbf{e}_{i,k}$ ($i=r,\theta,\phi$) are 
the $i$-th velocity component of the $k$-th star, and the basis unit
vectors of the spherical coordinate system at the position of the $k$-th
star, respectively. 

Given the assumptions:
 
\begin{itemize}
	\item (A1) There is no correlation between the velocity and position of a star
	\item (A2) The distributions of $v_r$, $v_\theta$, and
$v_{\phi}$ are symmetric around $v_r=0$, $v_\theta=0$, and 
$v_{\phi} = V_{\rm rot}$, respectively 
\end{itemize}

\noindent then the data points $\{ ( Q_{\phi,k}, v^{los}_k ) \}$ are likely to 
be distributed around a linear function of form: 

\begin{equation}
	v^{los} = V_{\rm rot} Q_{\phi} 
\end{equation}

\noindent Thus, by performing a linear fit to the data in the $Q_{\phi}-v^{los}$ plane, 
we can obtain $\hat{V}_{\rm rot}$ -- by which we denote the estimated
value of $V_{\rm rot}$ -- by measuring the slope of the best-fit linear
function. This approach is similar to that of \cite{FW1980}.

\subsection {Derivation of 3D Velocity Dispersion}

In estimating the velocity dispersion, we use a modified version of 
the approach used by \cite{Woolley1978}. 
In principle, this method utilizes the direction dependence of the line-of-sight velocity dispersion.

In addition to (A1) and (A2) above, let us further
assume that the velocity distribution of the sample stars satisfies the
following: 

\begin{itemize}
	\item (A3) The velocity ellipsoid is aligned with a spherical coordinate system 
	\item (A4) The dispersions in the distributions of $v_r$, $v_\theta$, and $v_{\phi}$ 
               around their centers are $\sigma_r$, $\sigma_\theta$, and $\sigma_\phi$, respectively 
\end{itemize}

\noindent Then, it can be shown, for $i=r,\theta,\phi$ 
(see \citealt{Morrison1990} for the case of $i=\phi$), that 
\begin{multline}
	{\rm E} \left[ {\rm var} \left[ v^{los} Q_{i} \right] \right] \\
		= \sigma_{r}^2 		\frac{1}{n} \sum_{k=1}^n Q_{i,k}^2 Q_{r,k}^2 
		+ \sigma_{\theta}^2 	\frac{1}{n} \sum_{k=1}^n Q_{i,k}^2 Q_{\theta,k}^2 
		+ \sigma_{\phi}^2 	\frac{1}{n} \sum_{k=1}^n Q_{i,k}^2 Q_{\phi,k}^2 \\
		+ \frac{V_{\rm rot}^2}{n-1} \left\{ \sum_{k=1}^n Q_{i,k}^2 Q_{\phi,k}^2	- \frac{1}{n} {\left( \sum_{k=1}^n Q_{i,k} Q_{\phi,k} \right)}^2 \right\}, 
\end{multline}

\noindent where we denote the expectation and variance of $X$ by ${\rm E}[X]$ and ${\rm var}[X]$,
respectively. 

Thus, by substituting the observed $\left( {\rm var} \left[ v^{los} Q_{i}
\right] \right)$ for ${\rm E} \left[ {\rm var} \left[ v^{los} Q_{i}
\right] \right]$, and the already estimated $\hat{V}_{\rm rot}$ for
$V_{\rm rot}$, we obtain:
\begin{multline} \label{matrix eq}
	\left[
		\begin{array}{c}
			{\rm var} \left[ v^{los} Q_{r}	\right] \\
			{\rm var} \left[ v^{los} Q_{\theta}	\right] \\
			{\rm var} \left[ v^{los} Q_{\phi}	\right] \\
		\end{array}
	\right]
	 - \frac{\hat{V}_{\rm rot}^2}{n-1}
	\left[
		\begin{array}{c}
			\sum_{k} Q_{r, k}^2       Q_{\phi, k}^2 	- \frac{1}{n} {\left( \sum_{k} Q_{r, k}       Q_{\phi, k} 	\right) }^2\\
			\sum_{k} Q_{\theta, k}^2 Q_{\phi, k}^2 	- \frac{1}{n} {\left( \sum_{k} Q_{\theta, k} Q_{\phi, k} 	\right) }^2\\
			\sum_{k} Q_{\phi, k}^4 			- \frac{1}{n} {\left( \sum_{k} Q_{\phi, k}^2			\right) }^2\\
		\end{array}
	\right] \\
	= \frac{1}{n} 
	\left[
		\begin{array}{ccc}
			\sum_{k} Q_{r, k}^4 			& \sum_{k} Q_{r, k}^2       Q_{\theta, k}^2 	& \sum_{k} Q_{r, k}^2       Q_{\phi, k}^2 	\\
			\sum_{k} Q_{r, k}^2 Q_{\theta, k}^2 	& \sum_{k} Q_{\theta, k}^4 			& \sum_{k} Q_{\theta, k}^2 Q_{\phi, k}^2 	\\
			\sum_{k} Q_{r, k}^2 Q_{\phi, k}^2 	& \sum_{k} Q_{\theta, k}^2 Q_{\phi, k}^2	& \sum_{k} Q_{\phi, k}^4 			\\
		\end{array}
	\right] 
	\left[
		\begin{array}{c}
			\hat{\sigma}_{r}^2 \\
			\hat{\sigma}_{\theta}^2 \\
			\hat{\sigma}_{\phi}^2 
		\end{array}
	\right] . 
\end{multline}

\noindent The solutions $(\hat{\sigma}_r^2, \hat{\sigma}_\theta^2, \hat{\sigma}_\phi^2)$ 
for this equation are the unbiased estimates for $(\sigma_r^2, \sigma_\theta^2,
\sigma_\phi^2)$. 
Note that these estimated values are not guaranteed to
take on positive values, so that we have to confirm the robustness of our
method via an independent Monte Carlo simulation, as described below. 

\subsection{Testing the Reliability of our Method}

In order to estimate the error that accompanies our estimate of the
three-dimensional velocity dispersion, we construct mock catalogs, and
perform simulated observation of mock stars drawn from these catalogs.
In this Monte Carlo simulation, each mock catalog is designed so that
the distribution of $r$ for the relatively metal-rich (or very
metal-poor) mock stars resemble those of the observed relatively
metal-rich (or very metal-poor) BHB stars, and that the velocity of the
mock stars obey a given anisotropic Gaussian velocity distribution (with
or without net rotation). We vary the velocity anisotropy parameter,
$\beta = 1 - (\sigma_\theta^2 + \sigma_\phi^2)/(2 \sigma_r^2)$, and
produce 1000 mock catalogs for each case. We find that the derived
$\sigma_\theta$ and $\sigma_\phi$ is only reliable for $r<16\; {\rm
kpc}$ and $r<18\;{\rm kpc}$ for the relatively metal-rich and very
metal-poor sample, respectively, while estimates of $\sigma_r$ are
reliable at any $r$.

\section{Application of our Methodology -- Derivation of Rotation Velocities, Velocity
Dispersions, and Velocity Anisotropy Parameters}

Throughout this paper, we assume that the Local Standard of Rest (LSR)
is on a circular orbit with a rotation speed of $220\;{\rm km\;s^{-1}}$
\citep{Kerr1986}. It is worth noting that our assumed values for the
Galactocentric distance of the Sun, $R_{\odot} = 8.5\;{\rm kpc}$, and
the circular velocity of the LSR are both consistent with two recent
independent determinations of these quantities by \cite{Ghez2008} and
\cite{Koposov2010}. \cite{Bovy2012} have recently determined, on
the basis of accurate line-of-sight velocities for stars in the APOGEE
sub-survey of SDSS-III, that the circular velocity of the LSR is close
to $220\;{\rm km\;s^{-1}}$. We also assume that the peculiar motion of the Sun
with respect to the LSR is $(U_\odot, V_\odot, W_\odot) = (10.0, 5.3,
7.2) \;{\rm km\;s^{-1}}$ \citep{Dehnen1998}. 

\begin{figure}
\begin{center}
	\includegraphics[angle=-90,width=0.8\columnwidth]{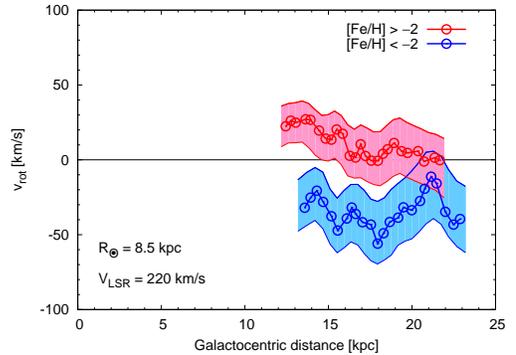} 
\end{center}
\caption{
The mean rotational velocity, $V_{\rm rot}$, for the relatively
metal-rich BHB sample (red) and very metal-poor BHB samples
(blue), as a function of Galactocentric distance $r$. Open
circles show the result for our BHB samples. In this plot, we bin the
relatively metal-rich (and very metal-poor) BHB stars in $r$ by binning
500 stars (400 stars), sorted in $r$, and moving through the sample in
steps of 20 stars. Each bin contains stars with a typical standard
deviation in $r$ of $\sim 3\;{\rm kpc}$, and the resultant $V_{\rm rot}$
is presented at the median value of $r$. The associated shaded regions
represent the uncertainties of our results, estimated from the bootstrap
method, and denote the range covered by the 16\% and 84\% percentiles. }
\label{fig1}
\end{figure}

Figure \ref{fig1} shows the determination of mean rotational velocity,
$V_{\rm rot}$, for the relatively metal-rich (red) and very metal-poor
(blue) BHB samples, as a function of Galactocentric distance, $r$. The
shaded regions indicate the uncertainties in each result, estimated from
a bootstrap approach (sampling with replacement). Inspection of this
figure suggests that $V_{\rm rot}$ at $13-23\;{\rm kpc}$ is a slightly
decreasing function of $r$ for the relatively metal-rich sample, while
that for the very metal-poor sample is more or less flat. The relatively
more metal-rich sample is in modest prograde rotation, with $V_{\rm
rot}\sim 0-20\;{\rm km\;s^{-1}}$, while the very metal-poor sample is in
retrograde motion, with $V_{\rm rot}\sim -20\;{\rm to}\;-50\;{\rm km\;
s^{-1}}$, and lags that of the relatively metal-rich sample by $\sim
20-50\;{\rm km\;s^{-1}}$, similar to previous observations of stars in a
much more local sample by \cite{Carollo2007, Carollo2010}, the
outer-halo analysis based on distribution-function fitting
\citep{Deason2011}, and some recently simulated stellar haloes (e.g.,
\citealt{Tissera2012}). It is also intriguing to see that the rotational
shear between two samples seems to shrink at $r\sim 20\;{\rm kpc}$. 
We confirm that varying the LSR velocity only changes the absolute 
value of $V_{\rm rot}$, and that the very metal-poor sample {\it always} 
lags behind the relatively metal-rich sample, independent of the assumed 
LSR velocity.

\begin{figure}
\begin{center}
	\includegraphics[angle=-90,width=0.8\columnwidth]{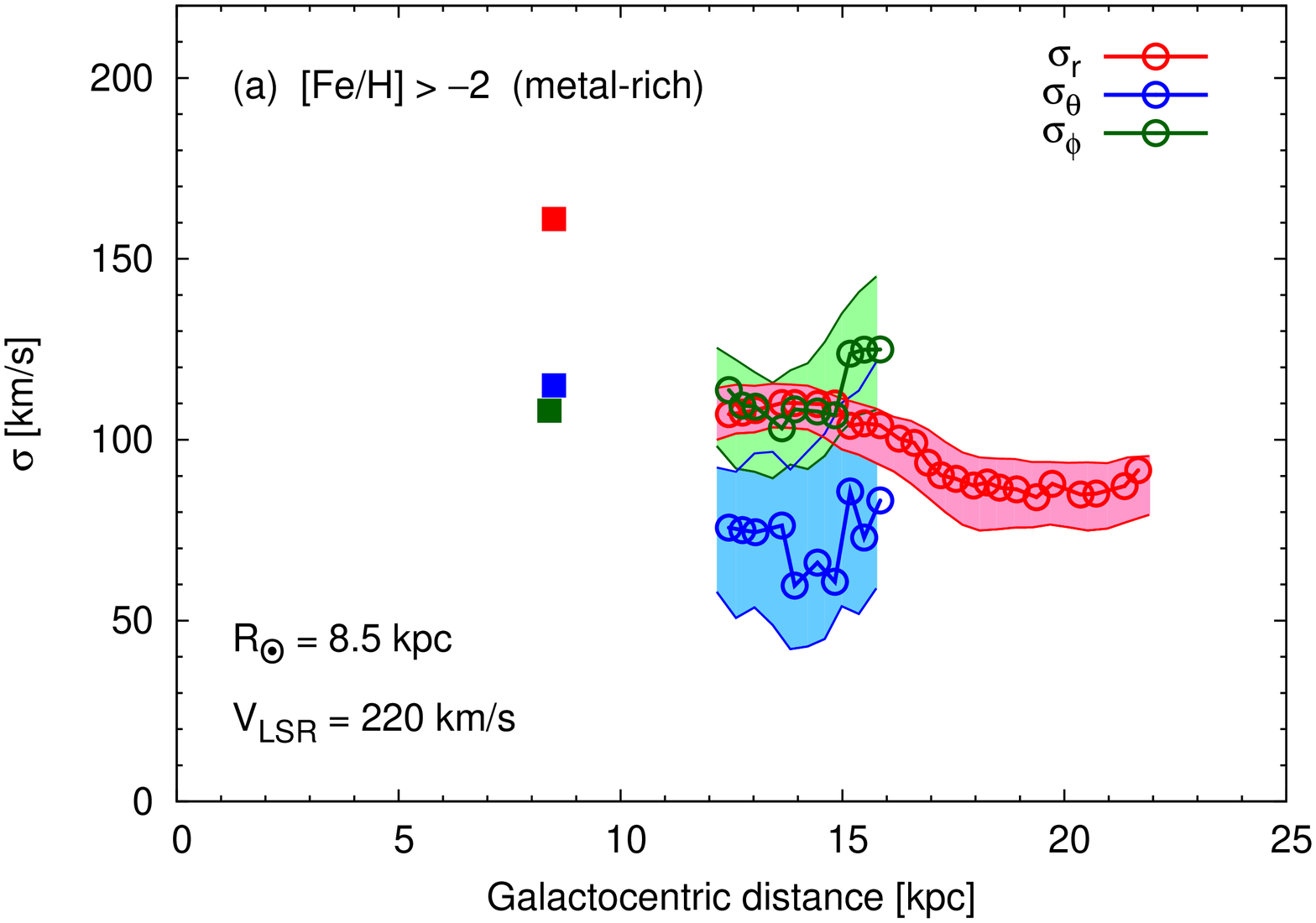} 
	\includegraphics[angle=-90,width=0.8\columnwidth]{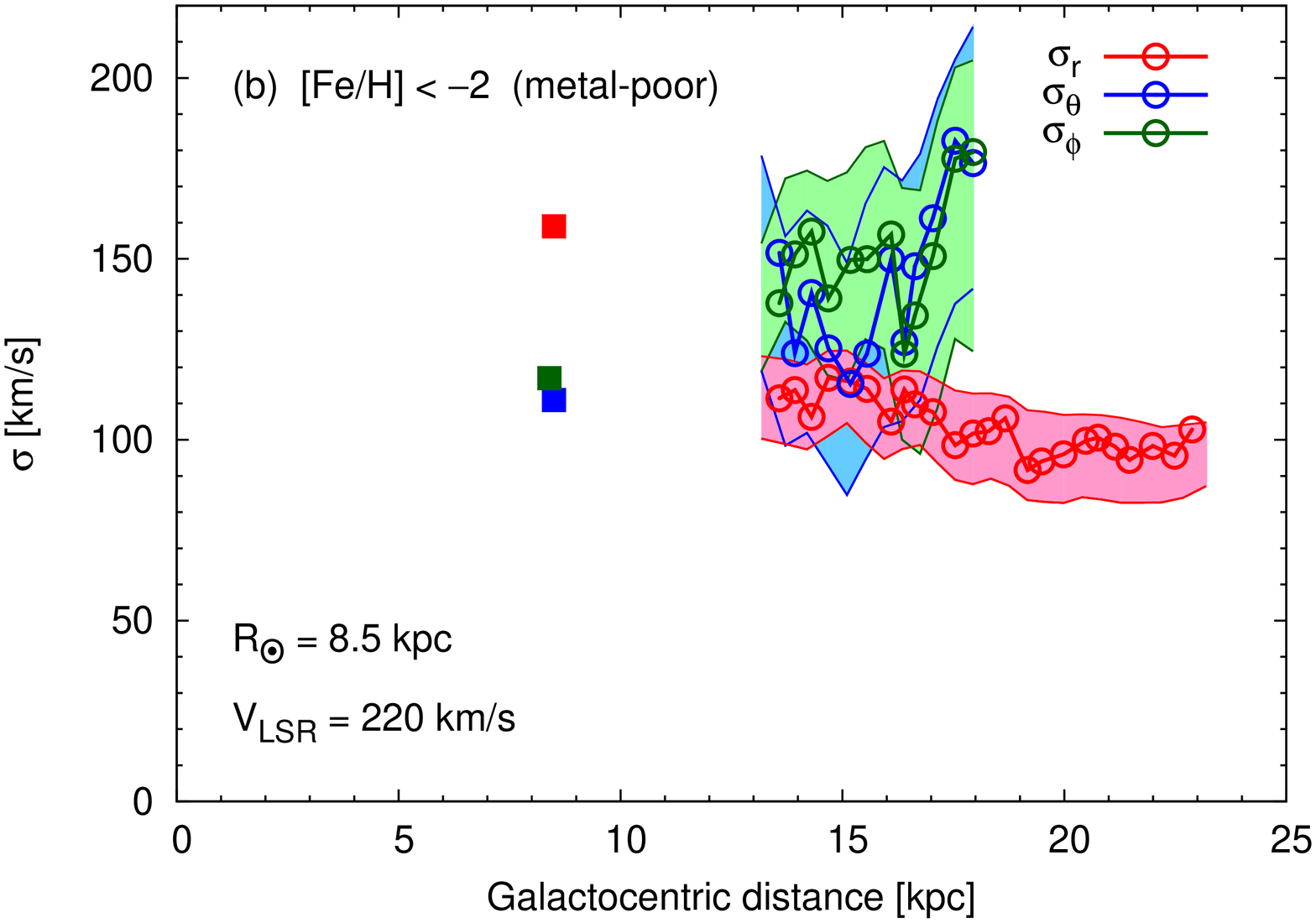} 
\end{center}
\caption{
Three-dimensional velocity dispersions $\sigma_r$ (red), $\sigma_\theta$
(blue), and $\sigma_\phi$ (green), as a function of Galactocentric
distance. Open circles show the result for our BHB samples. The same
binning procedure as in Figure \ref{fig1} is adopted. For the tangential
velocity dispersions, some of the bins with large Galactocentric distance
are excluded, because of a large systematic error indicated from Monte
Carlo simulations. The associated shaded regions represent the
uncertainties of our results, estimated from the bootstrap method. Results for the
relatively metal-rich BHB sample and very metal-poor BHB sample are shown in panel (a)
and (b), respectively. Filled squares at $r=8.5\;{\rm kpc}$ in panel (a)
and (b) represent the Solar-neighborhood observations of halo stars with
$-2<{\rm[Fe/H]}<-1.6$ and ${\rm[Fe/H]}<-2$, respectively, taken from \cite{CY1998}.}
\label{fig2}
\end{figure}

Figure \ref{fig2} shows determinations of the radial velocity
dispersion, $\sigma_r$, and the two components of the tangential
velocity dispersion, $\sigma_\theta$ and $\sigma_\phi$, for the
relatively metal-rich and very metal-poor samples, as a function of $r$.
As seen in the figure, $\sigma_r$ is about $100\;{\rm km\;s^{-1}}$, and
exhibits a declining behavior over $12.5 < r < 20\;{\rm kpc}$ for both
samples, supporting most previous studies (e.g., \citealt{Xue2008,
Brown2010}). In addition, this figure suggests that $\sigma_\theta
\simeq \sigma_\phi$ holds for the very metal-poor sample. Noting that
the gravitational potential is nearly spherical well above the disk
plane, and that $V_{\rm rot}$ of the very metal-poor stars is small
compared with its velocity dispersion, it follows that they obey a
distribution function that depends mainly on the orbital energy and
angular momentum \citep{BT2008}, and thus their spatial distribution is
nearly spherical. The different behavior of the 3D velocity dispersions
for the two samples implies that both the functional form of the
distribution function and spatial distribution for these samples are
different. 

\begin{figure}
\begin{center}
	\includegraphics[angle=-90,width=0.8\columnwidth]{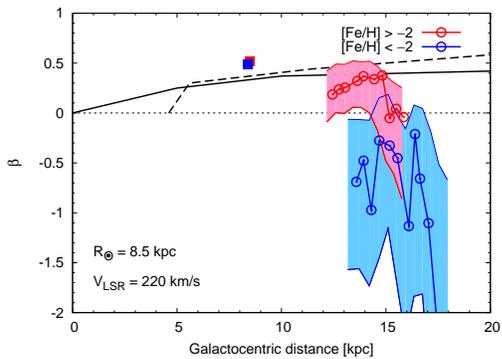} 
\end{center}
\caption{
The velocity anisotropy parameter, $\beta = 1 - (\sigma_\theta^2 +
\sigma_\phi^2) / (2 \sigma_r^2)$, for the relatively metal-rich (red)
and very metal-poor
(blue) BHB samples, as a function of Galactocentric distance. Open circles
show the result for our BHB samples, and correspond to those bins
presented in Figure \ref{fig2}. The associated shaded regions represent
the uncertainties of our results, estimated from the bootstrap method.
Filled squares at $r=8.5\;{\rm kpc}$ represent the Solar-neighborhood
observations of halo stars with $-2<{\rm[Fe/H]}<-1.6$ (red) and ${\rm[Fe/H]}<-2$
(blue), taken from \cite{CY1998}. The black solid
and black dashed line represent the $\beta$ profiles of simulated
stellar haloes from \cite{Diemand2005} and \cite{Sales2007},
respectively. 
The dotted horizontal line at $\beta=0$ is added to guide the eye. }
\label{fig3}
\end{figure}

Figure \ref{fig3} shows the velocity anisotropy parameter, $\beta = 1 -
(\sigma_\theta^2 + \sigma_\phi^2)/(2 \sigma_r^2)$. This parameter
quantifies the relative dominance of the radial and tangential velocity
dispersions, and provides a simple diagnostic of the orbital properties
of our sample stars. If $\beta$ takes on values $\beta<0$, we can infer
that the halo stars are dominated by round orbits, while if $0<\beta<1$,
we can infer that the halo stars are dominated by radial orbits. When
$\beta$ is (nearly) zero, we can infer that the velocity distribution is
(nearly) isotropic. Figure \ref{fig3} suggests that relatively
metal-rich BHB stars are dominated by radial orbits at $12 < r/{\rm kpc}
< 15$ ($\beta = 0.3 \pm 0.3$), while the very metal-poor BHB stars are
dominated by circular orbits over $13 < r/{\rm kpc} < 18$ ($\beta = -0.9
\pm 0.7$). A similar trend in the eccentricity distribution is
also confirmed in the more local sample of halo stars reported by
\cite{Carollo2010}, in which very metal-poor halo stars possess a higher
fraction of low-eccentricity orbits (see their Figure 5). 

\section{Discussion and Conclusions}

The observed systematic differences in the orbital motions of the
relatively metal-rich and very metal-poor BHB stars suggest that the
orbital motion of the progenitor systems of the stellar halo depends on
their metal abundances. Noting that dwarf galaxies with smaller total
stellar masses tend to have lower metal abundances (\citealt{Kirby2008},
and references therein), which can be understood as a result of a
smaller rate of metal-enrichment events such as
supernova explosions, our findings suggest that low-gas-fraction
(`star-rich') systems tend to move in radial orbits, while
high-gas-fraction (`gas-rich') systems tend to move in round orbits when
they convert their gas to stars. 

In the hierarchical galaxy formation scenario, the Milky Way (and other
large systems) attains its mass as a result of the mergers of infalling
smaller systems \citep{White1978, Blumenthal1984}. Due to the deep
gravitational potential well of the Milky Way, these infalling systems
tend to have radial orbits \citep{Sales2007}. At this stage, we expect
that the orbital properties of infalling systems is independent of
whether they are star-rich or gas-rich. However, the situation may be
different when such radially infalling systems pass near the Galactic
center. 
In such a region, star-rich systems 
(e.g., massive dwarf galaxies with low gas fractions) 
are expected to be disrupted
by tidal interactions, which give rise to field halo stars with radial
orbits \citep{Sales2007}. On the other hand, gas-rich systems are
expected to interact with other gas-rich systems, and lose some orbital
energy via dissipational processes \citep{SL1989, Theis1996, Sharma2012}. 
If the angular momentum of such a gas-rich system is approximately conserved, the orbit is circularized, 
and the pericentric distance (distance of closest approach
to the Galactic center) increases. Once its pericentric distance become
sufficiently large to avoid the central region of the Milky Way, further
orbital change becomes less likely, since encounters with other gas-rich
systems becomes less probable. Therefore, gas-rich systems tend to move
in round orbits. When the Galactic disk forms, some gas-rich systems
with similar orbital motions to disk gas may be absorbed into the disk,
due to their small relative velocities, while others remain moving on
orbits across the halo, and eventually form field halo stars with round
orbits. If the total orbital angular momentum of gaseous systems is
initially near zero, these halo stars would exhibit a net retrograde
rotation with respect to disk stars. 
Some gas-rich systems in the Milky Way that 
have not yet formed many stars 
might be associated with the observed high-velocity clouds \citep{Blitz1999, Putman2012}.

In the Solar neighborhood, observations suggest that $\beta \simeq 0.5$ for
halo stars, almost independent of metal abundance 
(\citealt{CY1998}; see also \citealt{Yoshii1979}), as 
overplotted in Figure \ref{fig3}. Combined with our result, we see that
the motion of the relatively metal-rich BHB stars move on radial orbits
at $8.5 < r/{\rm kpc} < 15$. This behavior is consistent with recent
cosmological simulations in which most halo stars originate from
accreted dwarf galaxies \citep{Diemand2005, Sales2007}, as also
overplotted in Figure \ref{fig3}. However, our very metal-poor halo BHB
stars suggest a transition from radially-anisotropic to
tangentially-anisotropic velocity distributions. The existence of very
metal-poor outer-halo stars with round orbits, which is not confirmed in
simulated stellar haloes, suggests that current simulations of disk
galaxy formation may lack some important mechanisms, such as those
proposed in the previous paragraph. We here note that many authors have
pointed out that the simulated dwarf galaxies show an over-production of
stars, when compared with observed dwarf galaxies \citep{Sawala2011}.
This phenomenon is often called the ``overcooling problem'', and this
might result in underestimation of the gas-rich progenitors in simulated
haloes, which in turn underestimate the numbers of metal-poor halo stars with round
orbits.  

We find that the kinematics of outer-halo stars exhibit a marked
dependence on stellar metal abundance, which provides information about the
physical properties of their progenitor systems (such as the gas fraction). 
Both of our main results, that the relatively metal-rich and
very metal-poor stars are dominated by radial and round orbits,
respectively, and that the mean rotational velocity of very metal-poor
halo stars lags that of relatively metal-rich halo stars, can be
explained if very metal-poor stars originate from gas-rich systems and
metal-rich stars from star-rich systems. Our findings cast a new light 
on the formation mechanism of the Milky Way and similar disk galaxies.

\acknowledgments{

We thank Naoto Kobayashi, Tsuyoshi Sakamoto, and Shigeki Inoue for
useful discussions. KH is supported by JSPS Research Fellowship for
Young Scientists (23$\cdot$954). TCB and YSL acknowledge partial support for
this work from PHY 02-16783 and PHY 08-22648: Physics Frontiers Center /
JINA, awarded by the US National Science Foundation. }

\end{document}